# The case for balanced hypothesis tests and equal-tailed confidence intervals

André GILLIBERT[ab][*][†], Jacques BÉNICHOU[bc] and Bruno FALISSARD[a]

[a] INSERM UMR 1178, Université Paris Sud, Maison de Solenn, Paris, France.

[b] Department of Biostatistics and Clinical Research, CHU Rouen, Rouen, F 76031, France

[c] Inserm U 1181, Normandie University, Rouen, France

[*] Correspondence to: André GILLIBERT, Department of Biostatistics and Clinical Research, CHU Rouen, Rouen, F 76031, France

[†] E-mail: andre.gillibert@chu-rouen.fr

# 1 Abstract

**Introduction**: there is an ongoing debate about directional inference of two-sided hypothesis tests for which some authors argue that rejecting $\theta = \theta_0$ does not allow to conclude that $\theta > \theta_0$ or $\theta < \theta_0$ but only that $\theta \neq \theta_0$, while others argue that this is a minor error without practical consequence.

**Discussion**: new elements are brought to the debate. It is shown that the directional interpretation of some non-directional hypothesis tests about Receiver Operating Characteristic (ROC) and survival curves may lead to inflated type III error rates with a probability of concluding that a difference exists in the opposite side of the actual difference that can reach 50% in the worst case. Some of the issues of directional tests also apply to two-sided confidence intervals (CIs). It is shown that equal-tailed CIs should be preferred to shortest CIs. New assessment criteria of two-sided CIs and hypothesis tests are proposed to provide a reliable directional interpretation: partial left-sided and right-sided $\alpha$ error rates for hypothesis tests, probabilities of overestimation and underestimation $\alpha_L$ and $\alpha_U$ and interval half-widths for two-sided CIs.

**Conclusion**: two-sided CIs and two-sided tests are interpreted directionally. This implies that directional interpretation be taken in account in the development and evaluation of confidence intervals and tests.

# 2 Introduction

Hypothesis tests may be performed on a statistic $\theta$, such as the effect of a treatment or an exposition. For a prespecified $\theta_0$, a two-sided hypothesis test is typically based on the rejection of the null hypothesis that $\theta = \theta_0$ and allows the conclusion of the alternative hypothesis that $\theta \neq \theta_0$.

Uncertainty about an estimation of the statistic $\theta$ may also be expressed as confidence intervals (CIs). The most used CIs are two-sided with a confidence level set at $1 - \alpha$, typically 95%. Two-sided CIs procedures are designed to guarantee that the probability $\Pr(\theta \in CI)$ that the CI contain the statistic $\theta$ be as close as possible as the confidence level $1 - \alpha$.

Although hypothesis tests and CIs have different definitions, they are related. For a given hypothesis test procedure, the set of $\theta_0$ values that are not rejected by the hypothesis test is a CI. For a given CI, the hypothesis that $\theta = \theta_0$ can be rejected when the CI does not contain $\theta_0$.

There is an ongoing debate about whether two-sided hypothesis tests should be used or replaced by one-sided, directional or three-sided hypothesis tests [1–4]. The literature about directional inference on CIs is scarce but similar theoretical considerations apply.

First, we will summarize the arguments for and against the use of two-sided tests. Second, we will see that proponents of both positions agree on how statistics should be interpreted but they disagree on the severity of the consequences of the use of two-sided tests that were not designed for directional inference. Third, we will show that the consequences of improper interpretation of two-sided inference can be more severe that had been previously reported. Fourth, we will show that these considerations should be taken in account when designing and assessing new and old statistical tools. Fifth, we will draw a parallel between hypothesis tests and CIs to transfer hypothesis tests considerations to CIs.

# 3  Arguments for and against two-sided hypothesis tests

Kaiser [5] argues that rejecting the hypothesis $\theta = \theta_0$ should lead to the conclusion that $\theta \neq \theta_0$ but not that $\theta > \theta_0$ or that $\theta < \theta_0$ according to the sign of the observed difference [5]. He proposes a directional theoretical framework to permit directional conclusions, with three hypotheses:

H1 : $\theta < \theta_0$

H2 : $\theta = \theta_0$

H3 : $\theta > \theta_0$

This allows Kaiser to define the error $\gamma$ (type III error) of concluding about the existence of an effect in the opposite direction of the actual effect. In the classical two-sided framework, this error does not exist, because rejecting $\theta = \theta_0$ with $\hat{\theta} < \theta_0$ (observed value less than the test target) while $\theta > \theta_0$ (real value greater than the test target) is a correct conclusion rather than an error.

John E. Overall argues that in regulated fields, such as drugs, the risk of error is taken only on one side [6]. Showing the superiority of the treatment to the placebo while it is equal or inferior may lead to the marketing of an inefficacious or harmful treatment (right-sided type I error). Showing that the treatment is significantly inferior to the placebo leads to the same conclusion than failing to show its superiority: in both cases, the treatment will not be marketed. Therefore, this risk should add to the type II error rate rather than the type I error rate.

Lombardi and Hurlbert [2] argue that one-sided tests are unable to detect effects opposite to what was expected. They argue that performing a second one-sided test in the opposite direction is possible but doubles the $\alpha$ error rate. Indeed, a balanced two-sided test with $\alpha = 0.05$ is equivalent to two one-sided tests with $\alpha = 0.025$. Lombardi and Hurlbert argue that Kaiser directional two-sided theory (H1, H2, H3) is practically equivalent to the classical two-sided theory (H0: $\theta = \theta_0$ vs H1: $\theta \neq \theta_0$) with directional conclusions based on whether $\hat{\theta} > \theta_0$ or $\hat{\theta} < \theta_0$ and that $\gamma$ (type III) errors are negligible.

Ludbrook [7] and Koch [8] suggest choosing the test on the *a priori* scientific hypothesis. The test alternative hypothesis is the scientific hypothesis performed before collecting data of the study and should be one-sided when a hypothesis is done in a specific direction. When no specific hypothesis is done, Ludbrook suggests that the test should be two-sided.

Ruxton and Neuhäuser [1] support two-sided tests but assume that their interpretation will be directional. They notice that the type I error may not split equally in two $\alpha/2$ errors but do not see that as a problem. Freedman [9] argues that, even if two one-sided tests are performed (one in each direction), the actual risk taken is not doubled compared to a single one-sided test. This can be understood with a simple example. If a comparison of two population means, $\mu_X$ and $\mu_Y$ must be performed with a single observation in each sample group, the one-sided test concluding that $\mu_X > \mu_Y$ if $m_X > m_Y$ has a one-sided type I error rate at 50%. It means that, if $\mu_X \leq \mu_Y$ (H0), in the worst case ($\mu_X = \mu_Y$) one out of two experiments will conclude that $\mu_X > \mu_Y$. A non-directional two-sided test concluding that $\mu_X \neq \mu_Y$ whenever $m_X \neq m_Y$ has a type I error rate equal to 100% if the distribution is continuous. A directional test concluding that $\mu_X > \mu_Y$ when $m_X > m_Y$ and that $\mu_X < \mu_Y$ when $m_X < m_Y$ has a type III error rate equal to 50% if one assumes that $\mu_X \neq \mu_Y$. Indeed, in the worst-case scenario where $\mu_X \approx \mu_Y$ but $\mu_X \neq \mu_Y$, the conclusion will be correct half of the time and wrong half of the time. Tukey argues that only the sign of the difference $\mu_X - \mu_Y$ is unknown and that the perfect equality $\mu_X = \mu_Y$ is never plausible [10]. The case where the difference $|\mu_X - \mu_Y|$ is negligible (equivalence) is another debate [4].

## 4 Summary of arguments

There is general agreement that conclusions should be directional, either with pre-specification of the direction (one-sided test), or with an *a posteriori* specification of the direction of the effect, according to the observed effect. The questions are about how to formulate the null and alternative hypotheses, whether to perform a classical two-sided test, one or two one-sided tests or more elaborate tests, and what type I error rate should be taken in each direction. As Freedman [9] showed, there is no need of summing the two one-sided error rates when performing a two-sided test, since the risk is only taken in the direction of the conclusion. Each error rate should be controlled separately, typically set at 2.5% each.

Until now, the debate focused on how tests should be used and interpreted, but not how they should be conceived and assessed. It was also assumed that the use of unbalanced two-sided test may, at worst, double the one-sided type I error rate, from $\alpha/2$ to $\alpha$. We will show that interpreting some two-sided hypothesis tests directionally can raise the one-sided $\alpha$ and even $\gamma$ error rates to 50%. We will show that some statistical tools have been developed to gain statistical power or precision while, at the same time, making directional inference impossible or highly biased.

# 5   Directional interpretations with risk of error up to 50%

## 5.1   Venkatraman's test

Venkatraman's test [11] compares receiver operating characteristic (ROC) curves. The null hypothesis of this test is that the two ROC curves are perfectly superimposed, which implies that the two areas under curves (AUC) are equal. It is based on a statistic summing absolute values of differences between the two ROC curves. When the two curves cross, it may have a high power to detect that they are not superimposed but gives no indication on which one has the greater AUC. Figure 1 shows crossing ROC curves from a simulated dataset, based on shifted log-normal distributions, where AUCs are close to each other. The dataset contains 200 observations, 100 with positive diagnostic and 100 with negative diagnostic. The real AUCs are 0.763, for the solid curve and 0.759 for the dashed curve, but the observed difference of AUCs is opposite to the real difference. Venkatraman's test is significant at the 5% threshold (p=0.03). A naïve interpretation of Venkatraman's test as a directional two-sided test comparing AUCs leads to a type III error, *i.e.* concluding that a difference exists in the opposite of the real direction. With 1000 Monte Carlo simulations the type III error rate ($\gamma$ risk) was estimated at 44%, while the statistical power was estimated at 50% and type II error rate ($\beta$ risk) at 6%. Since Venkatraman's test compares ROC curves rather than AUCs, it could be argued that this is a misinterpretation of Venkatraman's test. That is true, but how frequent is this misinterpretation?

A review of the 133 documents citing Venkatraman's article found by Google® Scholar on March 18, 2021, show that Venkatraman's test is mostly interpreted as a directional test comparing AUCs of ROC curves. After exclusion of duplicates (n=8), methodological articles (n=29), documents not published in scientific journals (n=31) and studies not actually using Venkatraman's test on real-life data (n=19), 46 articles were analyzed. A total of 32 (69.2%) articles improperly interpreted at least one significant Venkatraman's test as evidence of a higher actual AUC for the measurement having a higher observed

AUC, 7 (15.2%) interpreted non-significant differences as evidence of the equivalence of AUCs (n=4) or ROC curves (n=3) and 7 (15.2%) described AUCs (n=5) or ROC curves (n=2) as non-significantly different without further precision. In one study [12] the probability of a type III error was very high (~50%). Salcedo *et al* assessed the diagnostic AUC for 63 patients *vs* 622, with a caregiver symptom severity scale (CSS) and a caregiver symptom count scale (CSC). The CSS and CSC ROC curves crossed (see Figure 1 of Salcedo *et al* [12]). The CSS and CSC AUCs were respectively estimated at 0.79 (95% CI: 0.75 to 0.83) and 0.78 (95% CI: 0.74 to 0.83). Due to ROC curves crossing, the Venkatraman's test was significant (p=0.001) and authors concluded that the CSS outperformed the CSC. Due to the construction of CSC and CSS, a high covariance between the two scales is expected, but the observed AUC difference is so small (0.01) that it is probably smaller than the random error. It is possible that the CSC AUC is actually slightly higher than the CSS even though the opposite has been observed on the sample, leading to a type III error.

In order to identify the planned usage of Venkatraman's test, the example provided by Venkatraman in his original publication was analyzed [11]. The difference of ROC curves being non-significant in the example, Venkatraman concluded that "*Both the tests do not reject the hypothesis that the ROC curves are equal, suggesting that the effect of volume is the same for both dose levels.*" Therefore, this test is interpreted as an equivalence test by failed rejection of the null hypothesis, which is known to be an inappropriate interpretation [13].

## 5.2   Survival curves crossing

When Cox's proportional hazards assumption is violated, log-rank tests and Cox models are not recommended. Li *et al* reviewed 20 tests of comparison of survival curves having no proportional hazard assumptions [14], published in 14 articles; several tests with different tuning parameters could be analyzed for the same articles. All these tests were designed to reject the null hypothesis that the curves are perfectly superimposed. Fourteen (70%) tests published in ten articles gave no indication as to which curve is the "best" [15–24]. Indeed, the absolute values of differences or squared differences were summed or a maximum of differences between curves or hazard functions was calculated, making it impossible to identify the direction and position of the difference. Six (30%) tests published in four articles gave different weights to early and late events [25–28]; all were based on the log-rank or Kaplan-Meier with modified weights. Weights of Gehan-Breslow and Tarone-Ware are higher for early events than for late events. Although it may help to reject the null hypothesis of superimposed curves, it tends to favor the survival curve having a poor long-term outcome. Figure 2 shows survival curves crossing. Gehan-Breslow and Tarone-Ware when interpreted as directional tests from their Z statistic, lead to the conclusion that the curve with poor long-term survival (blue solid curve) has better overall survival since early events are much more weighted than late events. The log-rank test gives less weight to early events and concludes that the curve with good long-term survival (dashed red curve) has better survival. Therefore, tests designed to compare curves that cross, are either unable to provide a direction of the difference or tend to favor the curve with poor life expectancy. This is probably because most authors only assessed the statistical power to reject the null hypothesis of perfect superimposition of curves without directional interpretation in mind.

We performed a review of the literature that used the Gehan-Breslow test, with the keyword "Gehan[TW]" on PubMed, including all articles published from January 2000 up to March, 19$^{th}$ 2021. A total of 202 references were found. After exclusion of methodological articles (n=24), articles that did not apply any Gehan-Breslow test on any real-life dataset (n=36), and one article for which the full text could not be retrieved, the interpretation of the test was analyzed in 141 articles. A total of 68 of 141 (48.2%) articles interpreted the test as a comparison of an unrelated survival statistic, such as survival rate at a specific time (n=26), survival rate at time of last follow-up (n=6), median of survival (n=23), mean of survival (n=9) or hazard ratio (n=4). This was identified by sentences such as "*There was a median 48-month longer survival in patients with carboplatin HSR receiving carboplatin desensitization when compared to patients without carboplatin HSR (p = 0.0094)*". A total of 25 of 141 (17.7%) articles had a graphical interpretation of the difference suggested by sentences such as "*Kaplan-Meier survival curves showed significantly worse patient survival (figure 1) (p =0.004) and graft survival (figure 2) (p 0.02) for low T*". A total of 10 of 141 (7.1%) articles had a directional interpretation of the comparison without any specification of the statistic compared, with non-specific terms such as "*better survival*" or "*worse survival*" while 26 of 141 (18.4%) articles had a non-specific non-directional interpretation of the statistic such as "*When comparing the survival rates related to the alloy used, the Gehan-Wilcoxon test showed no significant differences*" and 8 of 141 (5.7%) provided no interpretation or an unclear interpretation, often due to the use of several different P-values (log-rank, Gehan-Breslow, etc.) or presentation of different statistics (median survival time, crude survival rate, etc.) in the same sentence. Finally, 4 of 141 (2.8%) articles interpreted the test as a comparison of early survival with sentences such as "*statistically significant delay of TRAMP-PSA tumor growth at early time points (Gehan-Breslow test, p = 0.002)*".

Articles interpreting the Gehan-Breslow test as a comparison of an unrelated survival statistic (*e.g.* median survival time) may, in the worst case have a type III error rate ($\gamma$ risk) close to 50%. Indeed, if medians are almost equal in two groups, the observed difference of medians would be randomly positive or negative in half of the experiments, although the Gehan-Breslow test may be almost always significant on a large sample.

Authors may not be taken responsible of misinterpretations of the tests they developed. However, some of them fooled themselves. For instance, Fleming *et al* [24] developed a modified Kolmogorov-Smirnov test for comparison of survival curves, that do not provide any indication on the direction of the difference. They provided an example from a real-life clinical trial in bile duct cancer and wanted to "*test whether patients treated with RöRx+5-FU would survive longer than control patients*" with their own procedure. Their test was able to show that curves were not superimposed but provided a directional interpretation of their test. Shen and Cai also provided a directional interpretation of a non-directional test in the example illustrating their article [20]. The absence of proper directional interpretation is less obvious for the test of Shen and Cai than for other tests because it can be used as a "one-sided" test trying to show evidence that a curve is better than the other, by choosing an optimal weight function, either favoring early or late events, to prove the hypothesis in that specific direction. Unfortunately, sampling fluctuations are computed under the null hypothesis of superimposed curves so that in cases of blatantly crossing curves, the test could show simultaneously the one-sided superiority of the first curve to the second (*e.g.* by choosing heavy early weights) and the superiority of the second curve to the first (*e.g.* by choosing heavy late weights). Indeed, according to the direction indicated by the statistician, the weights are automatically chosen by the procedure to prove what he wants.

From the 10 articles about tests with no possible directional interpretation cited in the systematic review of Li *et al* [14], 8 (80%) gave examples about trials in human or animal [16–21,23,24] for which a directional interpretation is expected, 1 (10%) [22] gave an example about a prognostic study for which a directional interpretation is expected and 1 (10%) [15] gave no example. From these 10 articles, only the two articles mentioned in the previous paragraph [20,24] gave a directional interpretation of their test in the example, all other concluded only that the curves significantly differed.

# 6 Consequences of inadequacy between theory and practice

The theory is often based on a non-directional null hypothesis while in practice, tests are interpreted directionally. In the next paragraphs, the consequences of this inadequacy between theory and practice are analyzed.

## 6.1 Development of tests without proper directional interpretation

Proponents of two-sided tests agree that the interpretation of two-sided tests is directional. This is confirmed by the literature review about articles using Venkatraman's and Gehan-Breslow's tests. This directional interpretation is not taken in account in many articles developing new statistical methods, especially for the problem of comparison of survival curves that cross. Due to an inappropriate null hypothesis, authors may ignore the fact that the main problem of crossing survival curves is not to prove that curves differ but to define a relevant criterion to identify which one is the "best". Authors developing new methods can fool themselves and fool others, as the literature reviews in the previous sections has shown.

## 6.2 Improper assessment of tests properties

Outside of survival or ROC curves analyses, even with simple one-dimensional statistics, ignoring the directional interpretation of two-sided tests may lead to an inappropriate assessment of the properties of the statistical test. A two-sided hypothesis test is designed to control the $\alpha$ risk of concluding $\theta \neq \theta_0$ when $\theta = \theta_0$, but may not control separately the one-sided risks of concluding that $\theta > \theta_0$ and that $\theta < \theta_0$ when $\theta = \theta_0$. In the worst case, the two-sided test may be equivalent to a one-sided test in an unknown direction, with all the $\alpha$ risk on one side. This would lead to a doubling of the $\alpha$ risk to prove one alternative hypothesis (e.g. $\theta > \theta_0$) and major loss of power to prove the other alternative hypothesis (respectively $\theta < \theta_0$). For instance, a test used to assess the superiority of a treatment to another may either have a very low statistical power to show the superiority or a doubled $\alpha$ risk, because of the unpredictable imbalance of the two-sided test. This could be especially unfair when there is no reason to favor one treatment compared to the other, such as a comparison of the medical to the surgical treatment of lumbar disc herniation. An imbalanced test could unfairly favor one treatment. Therefore, since two-sided tests are used directionally, they should be assessed with that use in mind, with a balance of the $\alpha$ risk, separated in two partial left-sided and right-sided risks controlled at $\alpha/2$. A two-sided test should be equivalent to two one-sided tests in opposite directions. As Freedman pointed out [9], the actual risk taken when performing two one-sided tests, is the one-sided risk, hence $\alpha/2$ when performing a balanced two-sided hypothesis test. As many hypothesis tests are quite balanced (e.g. Wald's tests of coefficients in generalized linear models with large sample sizes), most analyses described in the literature as two-sided with directional conclusions, actually take an $\alpha/2$ (usually 2.5%) risk. Unfortunately, the two-sided theoretical framework encourages authors of new tests to assess the global two-sided $\alpha$ risk without control of its balance between the left and right side of the alternative hypothesis.

# 7 Considerations for confidence intervals

A two-sided CI estimator, at confidence level $1 - \alpha$, must be built so that the frequentist probability that the CI contains the actual parameter is $1 - \alpha$ and the probability that the CI does not contain the actual parameter is $\alpha$. This $\alpha$ is the sum of the overestimation risk ($\alpha_L$), the probability that the CI is above $\theta$, and the underestimation risk ($\alpha_U$), the probability that the CI is below $\theta$. Two-sided CI estimators may be equal tailed, so that the underestimation and overestimation risks are equal or almost equal (*i.e.* $\alpha_U \approx \alpha_L \approx \alpha/2$ ). Unequal-tailed estimators may control the overall coverage but may have very different underestimation (*e.g.* 0.005) and overestimation (*e.g.* 0.045) risks. Unequal tailed CIs may be intentionally built to shorten the CI, such as Zieliński's interval [29]. As Zieliński show, a shortest CI may be built from an equal-tailed CI by moving both CI bounds upwards or downwards, increasing one risk (overestimation or underestimation) and decreasing the other (respectively underestimation or overestimation). Equal-tailed CI may be interpreted as the intersection of two one-sided CI while unequal-tailed CI bounds should not be interpreted separately since the actual risk associated to each bound is not well controlled.

In a phase III drug randomized controlled trial, the treatment effect may be estimated by CI. When filling a new drug application, the treatment effect may be claimed to be equal or greater than the lower bound of the confidence interval. The actual error rate is the one-sided error associated to the lower bound ($\alpha_L$). For an equal tailed CI, this risk is $\alpha/2$ , usually 2.5%. For an unequal-tailed CI, the actual error rate is unknown, between 0 and $\alpha$, and so, may be up to 5% for a 95% CI. Similarly, claiming that the frequency of adverse effects is less than the upper boundary of a two-sided CI has an error rate that is not well controlled by unequal-tailed CIs. Overestimation and underestimation are not equivalent, and one error rate cannot be traded for the other. Moving down the lower bound of the CI (decreasing the probability of overestimation) should not allow moving down the upper bound (increasing the probability of underestimation) since both movements tend to favor the hypothesis of low frequency of adverse effects.

Directional interpretations are done in all clinical and epidemiological contexts: to claim that an epidemiological exposure is dangerous because its effect is higher than the lower bound of the 95% CI or to claim that the sensitivity of a screening test is above the lower bound of the 95% CI. Therefore, CIs should be developed and assessed with these directional interpretations in mind. Equal-tailed CIs should be preferred over shortest CIs that may, in the worst case, be one-sided CIs in unpredictable directions. The probabilities of overestimation ($\alpha_L$) and underestimation ($\alpha_U$) should be both assessed when assessing CIs, rather than their sum $\alpha_L + \alpha_U = \alpha$. The argument for the use of shortest CIs is that they have better statistical precision, but this is not necessarily true when interpreting CIs directionally. When assessing the efficacy of a treatment, the expectancy of the lower bound should be as high as possible while the risk of overestimation is kept controlled to $\alpha/2$. That would guarantee the maximal statistical power and chance of concluding that the efficacy is greater than any predefined threshold. The expectancy of the lower bound can be indirectly assessed by the CI half-width, equal to the expectancy of the difference between the point estimate and the lower bound of the CI. A shortest CI, constructed by moving both bounds of an equal-tailed CI in the same direction, such as Zieliński's interval [29] or the highest density probability Bayesian credible intervals with a non-informative prior [30], may increase a half-width (*e.g.* Left half-width) while decreasing the other (respectively right half-width) at the same time, in an unpredictable way. This may sometimes reduce the statistical power when the CI is interpreted as a hypothesis test. Therefore, rather than assessing the total width of CIs, we suggest that the two half-widths of CIs, with their associated $\alpha_L$ and $\alpha_U$ risks be assessed when analyzing the statistical performances of CIs. This is equivalent to analyzing a two-sided CI as the intersection of two one-sided CIs.

# 8  Conclusion

Two-sided hypothesis tests and two-sided CIs are mostly used with a directional interpretation, although the theoretical framework on which they are based does not allow this interpretation. This gap between theory and practice has no practical consequences for many common tests, such as the likelihood ratio test for generalized linear models, but can sometimes lead to severely biased inference, such as for tests designed to compare survival or ROC curves that cross.

Entirely changing the biomedical research practice is not easy, especially because the most common tools have no major bias. However, the development of new statistical tools and assessment of existing tools should be performed with the directional interpretation in mind. We propose to assess separately, the partial left-sided and right-sided type I error rates for hypothesis tests, to guarantee their balance. We propose to assess separately the underestimation and overestimation risk of CIs as well as left and right half-width of the CIs. When developing new tests that cannot have a directional interpretation, we suggest to keep in mind the risk that it may be misinterpreted.

# 9  Supplementary Material

Detailed information about articles included in the literature reviews of the Venkatraman and the Gehan-Breslow tests is available on the Open Science Framework at https://osf.io/zds8b/

Figure 1: Receiver Operating Curves (ROC) from simulated datasets for which a naïve interpretation of Venkatraman's test leads to a type III error.

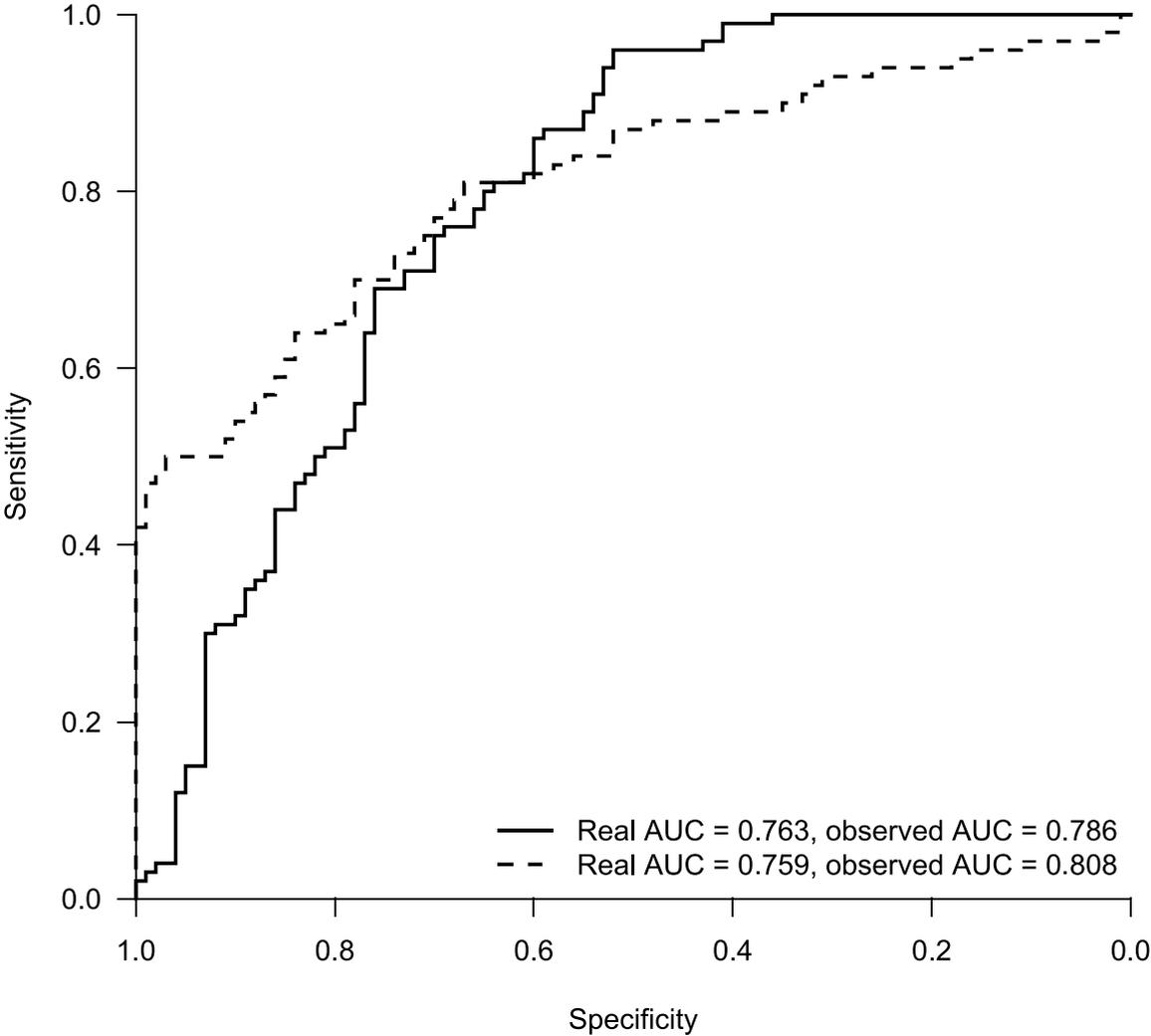

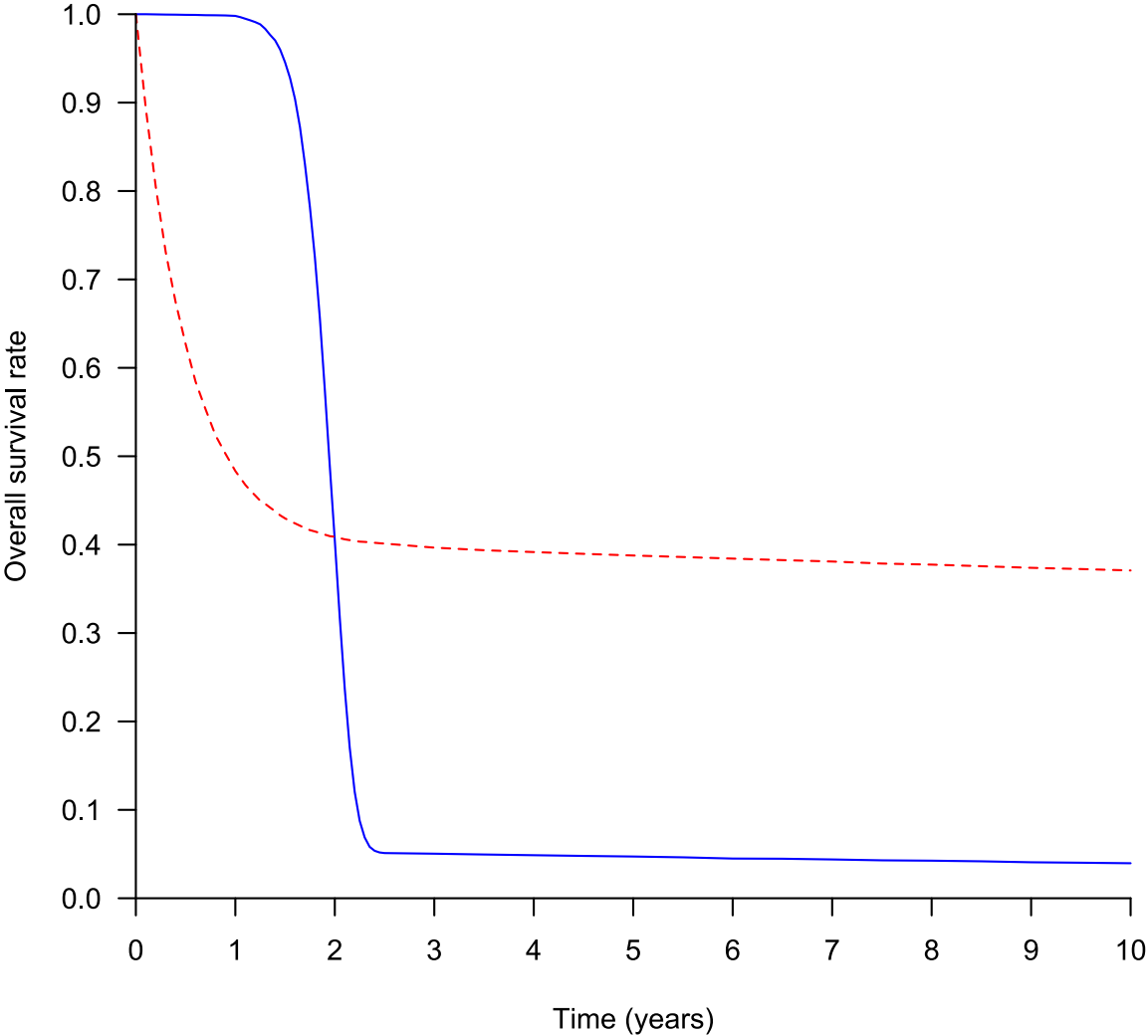

Figure 2: simulated survival curves crossing for which Gehan's test finds that the survival is better for subjects with poor long-term survival (blue solid line) than for subjects with poor short-term survival (red dashed line).